# Four-Step Approach Model of Inspection (FAMI) for Effective Defect Management in Software Development

[1]Suma V, [2]T. R. Gopalakrishnan Nair
[1]Member, Research and Industry Incubation Center, Bangalore, India.
sumavdsce@gmail.com
[2]Director, Research and Industry Incubation Center, Bangalore, India.
trgnair@ieee.org

**Abstract** – IT industry should inculcate effective defect management on a continual basis to deploy nearly a zero-defect product to their customers. Inspection is one of the most imperative and effective strategies of defect management. Nevertheless, existing defect management strategies in leading software industries are successful to deliver a maximum of 96% defect-free product. An empirical study of various projects across several service-based and product-based industries proves the above affirmations. This paper provides an enhanced approach of inspection through a Four-Step Approach Model of Inspection (FAMI). FAMI consists of i) integration of Inspection Life Cycle in V-model of software development, ii) implementation of process metric Depth of Inspection (DI), iii) implementation of people metric Inspection Performance Metric (IPM), iv) application of Bayesian probability approach for selection of appropriate values of inspection affecting parameters to achieve the desirable DI. The managers of software houses can make use of $P^2$ metric as a benchmarking tool for the projects in order to improve the in-house defect management process. Implementation of FAMI in software industries reflects a continual process improvement and leads to the development of nearly a zero-defect product through effective defect management.

## 1. INTRODUCTION

There are several factors, such as technological advances, high cost of software development and frequently changing business requirements, which influence the creation of high quality software. One of the challenging issues in software industry is to deliver high quality software to achieve total customer satisfaction. Effective defect management is one of the successful paths to produce high quality software. Two important approaches of effective defect management are quality control and quality assurance.

Testing is a quality control activity that addresses defects [1]. It can only show presence of defects and corrects them [2]. It is a defect detection activity [3] [4]. Inspection is one of the most effective formal evaluation techniques of quality assurance [5] [6] [7] [8] [9]. It detects most of the static defects at the early stages of software development and close to its origin [10]. It is both defect detection and prevention activity. Testing is constructively destructive while inspection is non-destructive activity [11]. Therefore, one of the successful managerial strategies to produce high quality product is to opt for both defect detection and defect prevention [12] [13] [14] [15].

Software industries follow several defect detection and prevention techniques to address defect related problems. Authors in [16] propose a new model in software development. Defect Prevention is one of the major benefits of this model. Authors in [17] describe 'when, who and how' approach for defect analysis. It enables to understand the quality control process and improvement areas. Author in [18] considers software quality problem as a major issue prevailing in industries. He suggests the need for improvement in existing quality strategies to address the issue.

Authors in [6] [19] express that inspection is the most mature, valuable and competent technique in the challenging area of defect management since three decades. However, the competitive tendency in effective defect management demands the development of research work towards enhanced approaches of inspection. Authors in [20] present a survey of research work in the area of inspection, since 1991 to 2005. They express that research in the direction of the introduction of novel inspection methods is very less. This motivated us to comprehend the existing inspection technique and thereby to introduce a novel and robust inspection model towards the development of nearly zero-defect product.

Despite the significance of inspection in effective defect management, many leading software industries continue to deploy products with residual defects. The aim of our research work was to determine the following:

- What is the outcome of existing inspection approach in defect management?
- Is there a need to introduce an enhanced approach for inspection technique to achieve effective defect management?
- Is there a need to measure the depth of effectiveness of inspection process?
- What is the extent of impact of inspection influencing parameters towards the effectiveness of inspection?
- Is there a need to introduce metrics to measure the inspection influencing parameters?
- Can Bayesian probability enable the management of software community to select the desirable ranges of inspection affecting parameters in order to achieve desirable and effective depth of inspection?

Organization of this paper is as follows. Section 2 contains the description of background elucidation for the inspection process. Section 3 explains the research methodology followed. Section 4 describes the empirical study of several projects from various leading service-based and product-based software industries. It further presents the observational inferences of the empirical data. Section 5 provides explanation of the enhanced inspection process.

Our Enhanced inspection process is a Four-Step Approach Model of Inspection (FAMI). It comprises of i) Integration of Inspection Life Cycle in V-model of Software Development Life Cycle ii) Implementation of process metric, 'Depth of Inspection (DI)' iii) Implementation of people metric, 'Inspection Performance Metric (IPM)' iv) Application of Bayesian probability for selection of appropriate values of inspection affecting parameters to achieve the desirable DI.

Implementation of FAMI reflects a continual process maturity in software industry. It further aims at development of nearly zero-defect product.

## 2. BACKGROUND ELUCIDATION FOR THE INSPECTION PROCESS

Author in [7] states the defect removal efficiency in the United States as on 2007 is up to 85%. This fact motivated us to analyze the effect of existing inspection technique in leading software industries towards effective defect detection.

Authors in [21] compare on several quality assurance models for small and medium-sized organizations. In addition, they in [22] propose a more realistic, flexible and cost effective inspection model for effective inspection process. This inspired us to analyze the existing inspection technique in small and medium scale industries under various categories of projects. The categorization includes very small projects with total development time of 250 person-hours to big projects of 10,000 person-hours.

Author in [23] emphasizes on estimating the defect type for the important deliverables to achieve effective defect removal. He expresses the significance of appropriate choice of defect detection technique and analyzes the factors influencing defect injection. This led us to analyze the percentage of occurrences of various defect types at major phases of development.

Authors in [24] recommend Evidence-Based Software Engineering (EBSE) approach for effective and efficient defect detection method. They suggest the use of inspection, for requirements and design phase and testing for code as the most beneficial defect detection technique. However, we emphasize on inspection at every phase of software development for effective defect management.

Authors in [25] suggest Defect Based Process Improvement. They use Defect Causal Analysis approach for process improvement through systematic reviews and learning through cause-effect relations. Authors in [9] emphasize upon inspection technique for effective defect removal. They express that meeting-based reviews do not have greater impact on defect capturing capability of review team than meeting-less methods. However, our study on empirical data clearly presents the impact of inspection preparation on inspection performance in industry atmosphere.

Author in [26] explores the effect of certain significant factors on the influence of inspection process in practice. A survey in Australia in 2003 explains the effect of implicit inputs (inspector characteristics) over explicit inputs (supporting documents). This encouraged us to analyze the parameters influencing inspection process. Further, the analysis enabled to introduce metrics to evaluate the effectiveness of inspection process.

Author [27] recommends the project managers to consider the peer review data for assessment of defect density, to reduce existence of latent defects and thereby enhance the quality of final product. Thus, our enhanced inspection model emphasizes upon peer review to be a formal and stringent activity of inspection.

Authors in [28] recommend the use of effectiveness of inspection team as a decisive measure for team size selection and as a tool for inspection planning. Our empirical investigation supports the suggestion of selection of desirable team size for effective inspection. However, our study further indicates that team size in isolation cannot provide greater impact on effective inspection process.

Authors in [29] accentuate upon implementation of requirements inspection in early phases of software development life cycle. However, we emphasize inspection at each phase of development for effective defect detection.

Authors in [30] put forth the necessity to reduce the residual defects. They recommend the formulation of strategies to reduce residual defects through creating awareness on defects in organizations, to re-evaluate the verification, validation and other such developmental policies to reduce residual defects. Implementation of our *$P^2$ metric* along with awareness of the defect pattern, which is exposed through our empirical study, aims to reduce the residual defects.

Author in [31] suggests the use of both inspection and testing to gain total customer satisfaction through high quality software. Hence, our enhanced inspection model emphasizes on both inspection and testing through integration of inspection life cycle in V-model, which emphasizes on both the strategies. Further, we recommend an *appropriate ratio* of inspection and testing effort during the software development.

Thus, i) State-of-the-art practice in defect management strategies, ii) requirement for continual quality improvement and iii) significance of inspection in effective defect detection in software development, focused us to study the existing defect management strategies in leading software industries.

The empirical analysis of data from the leading software industries enabled the comprehension of effectiveness of defect capturing capabilities through appropriate range of inspection time with testing time [32] [33]. Further, study in this direction, facilitated to analyze the possibility of occurrences of defect patterns at each phase of software development and to understand the defect severity levels.

Consequently, the observations made on the empirical data directed us towards the introduction of *$P^2$* quality metric for measuring the effectiveness of inspection.

FAMI is the result of the observations and inferences drawn from our empirical study of various projects across several software industries.

## 3. RESEARCH METHODOLOGY FOR EMPIRICAL STUDY

Our empirical study examines projects from leading service-based and product-based software industries.

*Challenges*
*Challenge 1:* Each company follows their self-defined strategy in development process.
*Challenge 2:* Generalization of several projects across various companies.
*Steps taken to overcome the challenges*
*Step towards Challenge 1:* All the companies are at CMM Level 5. Each company follows defect prevention activities and emphasizes on inspection technique.
*Step towards Challenge 2:* Unify the effort for comparison for each project against a standard measure. The standard measure that we have considered is total development time measured in person-hours.

The aim was to discover the effectiveness of existing inspection technique to deliver nearly zero-defect product.

*Hypothesis*

*Hypothesis 1:* The sampled projects are developed with similar technology, environment and programming language. They differ with respect to total development time measured in person-hour.

*Hypothesis 2:* Sampled data represent small, medium and big projects. Small project requires less than 1000 person-hour for its development. Medium projects need 1000 to 5000 person-hours of developmental time. Project, which demands more than 5000 person-hours for its development, is deemed to be a large project.

The above hypothesis in software industries helps to categorize projects for quantitative and metric collection purposes.

*Validity threat to the hypothesis*

*Validity threat to hypothesis 1:* Our observational inferences may not be applicable for projects developed with varying technology, environment and programming language.

*Validity threat to hypothesis 2:* Validity of the hypothesis may not be applicable in innovative projects with lack of domain knowledge on the new technology, environment and new programming language.

From the year 2000 onwards, many leading software industries started their journey towards implementation of CMM. Hence, scope of our analysis is related to projects developed from the year 2000 onwards to 2009.

The work began with empirical data collection of successful projects where effectiveness of inspection was visible. However, empirical data had few projects where inspection was ineffective, in successfully detecting defects. Hence, analysis of the successful nature of the projects was the first activity to perform. Verification of percentage of defects captured before the shipment of product against the total number of defects present in the product helped to comprehend the successful nature of the project.

The next step of work was to analyze the defect capturing capability of development team through existing inspection and testing approach. This required scrutiny of data at each phase of software development. To gain clarity over the work, only three major phases of software development namely requirement analysis, design and implementation phase are considered.

An analysis at this stage of work dealt with the scrutiny of defects and the time profile for the empirical projects at each phase. The knowledge of time scheduled for defect detection and awareness of defect pattern, demands an improvement in existing defect management strategies.

Further investigation enabled the analysis of factors that influence the effectiveness or ineffectiveness of inspection at every phase of software development. Hypothesis 1 through 2 enables to compare between two projects under each category. Their results led to the introduction of FAMI. The objective of FAMI is to deliver nearly zero-defect product to their customers.

## 4. OBSERVATIONS

Identified defects are categorized depending on defect type. They are blocker type of defect, which prevent continued functioning of the developer team, critical type of defect that results in software crash, system hang, loss of data etc.

Defect is categorized as a major type when a major feature collapses and a minor type when defect causes a minor loss of function, still allowing an easy work around. Trivial category of defect arises due to cosmetic problems. Based on these categories, severity levels are assigned as either urgent/show stopper, medium/work around or low/cosmetic [14]. Conversely, the authors in [34] would like to prioritize severity of defects based on demand driven computation using static profiling technique.

Nevertheless, in addition to the type of classification, it is vital to count the total number of defects. Further, the root cause analysis helps to analyze, eliminate and prevent the reoccurrences of defects.

An awareness of defect pattern enables to comprehend the severity of defect occurring at each phase of software development. It further enables to formulate defect management strategies on those process areas where defect is introduced.

Inspection greatly reduces defect propagation and hence its manifestation into the later stages of development [6]. Inspection process is highly influenced by nominal size of team, meeting effort and defect detection techniques adopted. These parameters further influences cost of rework by identifying different classes of defect severity [35].

Table 1 is a sampled data obtained from several leading service-based and product-based software industries. The sampled projects depict inspection time, preparation time, inspection team size and experience level of inspectors at each phase of development. In addition, they specify the type of defect. Furthermore, it indicates the successful nature of the project. The tables viz. Table 2, Table 3 and Table 4 depicts the percentage of time scheduled for inspection and the percentage of occurrences of various defect types at requirement analysis, design and implementation phase respectively.

### 4.1 Observational Inference

The empirical observation on the data from leading software industries shows the percentage of occurrence of various types of defects at each phase of development and for all the categories of projects. Tables 2 to 4 indicates the spanning of defect pattern for various types of defects. Table 5 depicts the existence of defect pattern across three major phases of development under the three major categories of projects. Observations from Table 5 indicate that possibility of percentage of defect occurrences at same phase for the sampled projects scales to same percentage. It indicates existence of defect pattern. This pattern is observed on empirical projects based on developmental time as a quality measure. In addition, the observations drawn from the Tables 2 through 4 are

- An inspection time of 10% at each phase of development results in 90% and above defect-free product.
- Projects P6 and P10 indicate that when defects captured through inspection at each phase of development is less than 30% then the total defect capturing capability of developer team is less than 90%.
- The existing inspection techniques in leading software industries show the capability of deploying an average of maximum of 96% defect-free product to their customers.

The above observations encouraged us to enhance the existing inspection technique (process) and accentuate upon inspection team (people) to deploy nearly zero-defect product (effective defect management).

Thus, our empirical observations indicate the existence of mutually cohesive triangle. Figure 1 depicts the mutually cohesive triangle. Process edge of the triangle demands enhancement to the existing inspection technique. People edge of the triangle emphasizes upon evaluation of inspection team who drive the process. Effective defect management, which is the tip of the triangle, depends upon the maturity level of process and people. The triangle, thus, stress upon the mutually dependent relation between process, people and effective defect management to generate high quality product.

FAMI is the result of implementation of mutually cohesive triangle.

## 5. INTRODUCTION OF FOUR-STEP APPROACH MODEL OF INSPECTION (FAMI)

Each phase of software development has deliverables. Each deliverable should undergo inspection for static defects and testing for dynamic defects. The aim is to detect defects before customer finds one. It ensures deployment of product with negligible risk. It builds confidence to ship the product. It further enhances quality of the product, increases productivity and reduces cost of rework [36] [10] [37] [38].

Inspection also detects the processes that initiate defects in the product. Table 1 indicates the capability of existing defect management techniques to deliver an average of maximum of 96% defect-free product. Therefore, the key challenge of progressive software industry is to deploy nearly zero-defect product. FAMI is an enhanced approach of software inspection that aims at effective defect management to achieve this.

### 5.1 FAMI STEP 1: Integration of Inspection Life Cycle in software development Process

As the first step, FAMI emphasizes upon the process improvement by integration of Inspection Life Cycle before Testing Life Cycle in V model approach of software development.

Figure 2 shows integration of Inspection Life Cycle in software development process. Inspection Life Cycle comprises of the following activities at every phase of development namely:
- Inspection of deliverables
- An inspection report
- Defect Trail
- Feedback on detected defects to the concerned authors for fixing them
- Re-inspection of deliverables
- Defect Trail (if any and report going back to authors)
- Consolidated Causal Analysis Report (CAR) and Non Conformance
- Report (NCR) closure report
- Defect Trail and NCR closure report to go to DP Process Centre
- Inspected deliverables to go to next phase of software development

During the requirements phase, the inspection artifact is the requirement specification. High-level design and low-level design are the inspection artifacts at design phase. Test case and source code are the inspection materials during implementation phase.

The outcome of inspection is an inspection report. It is a list of comments for the identified defects. It goes to the concerned author for rectification. Feedback mechanism facilitates developer team and management to identify and remove the defects along with fault processes. A detailed causal analysis for identified defects should be performed before final inspection of each deliverable.

Defect Trail maintains information about defects that includes type of defect, number of defects, root causes of defect, action items considered for rectification of the detected defects, inspectors involved and experience level of inspectors. This log is a lesson learnt for future development of the same project or for projects that are similar in nature.
Non Conformance Report (NCR) report includes quality audits performed as part of the organization's Quality Management System (QMS) requirements.

DP process centre maintains both defect trail and NCR closure report. DP process centre further discusses the preventive action for the mistakes committed and prepare plans accordingly for future work. The centre in addition keeps track of effectiveness of the suggested and implemented preventive action. Thus, inspection is both defect detection and prevention activity. It is also a constructive activity.

### 5.2 FAMI STEP 2: Introduction of Depth of Inspection (DI) as a Quality Metric

Effectiveness of inspection should be quantifiable [39]. Metrics acts as decision factors based upon which the inspection planning can be improved [40]. Second step of the model insists upon the need for implementation of quality metric to measure the effectiveness of inspection.

Total percentage of defects captured by inspection process is the ability of inspection process to capture defects. This definition indicates the depth in which inspection process has occurred. Accordingly, an operational definition for depth of inspection is the number of defects captured by inspection process from the total number of defects captured by both inspection and testing activities. The above operational definition enables to formulate a new process metric called 'Depth of Inspection (DI)'.

DI = Number of defects captured by inspection process (Ni) / Total number of defects captured by both inspection and testing (Td)

$$Di = Ni / Td \qquad (1)$$

Table 6 shows sampled fifteen projects that have given a solid indication that, value of DI metric is varying in the range of 0.21 to 0.67 at each phase of development. Thus, introduction of DI metric in industry indicates the success level of projects. Implementation of the metric indicates a refined development process [50].

Figure 3 shows the variation of DI curve with increasing complexity of the projects. DI value is a travel light indicating the depth in which inspection should carry on. The study of several projects reveals that DI value below 0.3 shows low level of efficiency of inspection. DI value in the range of 0.3 to 0.5 is normal efficiency of inspection. Table 6 proves the above argument. Competency of inspectors is necessary to attain DI value of 0.6 to 0.7. Figure 3 shows decreasing level of DI as the complexity of project increases irrespective of size of the project.

### 5.3 FAMI STEP 3: Parameters Influencing Depth of Inspection

Above observations presents a variation in DI value at each phase of software development. Third step towards enhanced inspection approach is to analyze all the parameters influencing DI. This focuses upon the people who drive the process.

Observations made across several projects have revealed the impact of certain parameters on DI. These parameters are inspection time, number of inspectors, experience level of inspectors, preparation (training) time and skill of inspectors. Figure 4 shows the impact of various parameters on Depth of inspection

Inspection time is a highly influencing parameter. Framework of time for inspection at all phases of development life cycle is necessary. Development of nearly zero-defect product is possible with an inspection time of 10%-15% out of total project time. An increase in inspection time reduces testing time [32].

Number of inspectors influences inspection process at each phase of development [41]. Figure 5 indicates the various activities performed within the inspection and re-inspection component of inspection life cycle. Self-reviewer, in compliance with self-review checklist will initially inspect each deliverable. Peer review is an Internal Quality Audit (IQA). It detects defects that have escaped from eyes of the author [42]. Hence, peer review is an effective defect detection and defect removal activity [43]. Technical leader or quality engineer performs External Quality Audit (EQA) for the deliverable. Identified defects further undergo causal analysis before final inspection. Project leaders and technical managers or senior managers perform final inspection. Project leaders responsible for the deliverable should not be the inspector for final inspection.

Figure 5 further shows the N weighted loop, which is a feedback report to the concerned author for fixing up of identified defects. The deliverables with rectified defects further undergo re-inspection before final inspection. The value of N weighted loop should be always less than two, to indicate the efficiency of inspection. Thus, for any project, it is important to have a minimum of three inspectors, which include self-inspector, peer reviewer and an external inspector.

Experience of inspectors is an influencing parameter in defect detection [44]. Experience level of inspectors can be considered in three categories. They are i) novice inspectors (up to 2 years) ii) average experienced inspectors ( above 2 years to 4 years) and iii) largely experienced inspectors (above 4). Hence, it is appreciable to involve inspectors with minimum of three years of experience with the domain knowledge.

Preparation (Training) time for inspector has a vital role in influencing the effectiveness of inspection [44][43]. Preparation time varies depending on the complexity of the project. An author of the deliverable conducts a walkthrough of the work product to the participating inspectors [38]. This enables inspection team members to analyze defects. Reading techniques further enhances the efficiency of inspector [45] [46] [47]. However, the whole effort of inspection can be further reduced using automation of inspection process [48]. Authors in [49] argue that number of inspectors does not affect efficiency of inspection meetings in defect detection. They suggest that inspection time and preparation time influences efficiency of inspection meeting. However, our empirical study reveals that inspection time, number of inspectors and their experience level with preparation time influences effective defect detection rate. Thus, it is vital to schedule minimum of 10% to 20% of total inspection time for preparation.

Thus, performance of an inspection process is the ability of an inspector to capture the maximum number of defects within the constraints of inspection influencing parameters. The formulation of above operational definition for inspection performance analysis introduces a new metric called Inspection Performance Metric (IPM) [51].

Inspection Performance Metric (IPM) = Number of defects captured by inspection process (Ni) divided by Inspection Effort (IE). Inspection effort (IE) = Product of Total number of inspectors (N) and Total amount of inspection time (T). And Total amount of inspection time (T) = Sum of Actual Inspection time (It) and Preparation time (Pt)

$$IPM = Ni / IE \qquad (2)$$

Where $IE = N \times T$ and $T = It + Pt$

Table 7 indicates the impact analysis of DI, IPM and experience level of inspectors over the sampled projects. Table 7 reflects the impact of IPM and experience level of inspectors on DI. Software organizations can choose an appropriate inspection effort according to their desired defect capturing maturity level.

### 5.4 FAMI STEP 4: Use of BBN for Selection of ppropriate Values of Inspection Affecting Parameters to Achieve the Desirable DI.

It is vital for management of software community to provide supportive environment for inspection activities [30]. We recommend management of software community to consider skill of inspectors and select desired ranges for the inspection parameters to achieve better results.

The fourth step of FAMI is application of Bayesian Belief Network (BBN). This enables to select appropriate values of inspection influencing parameters to achieve the desirable DI.

Bayesian approach helps to make decisions based on the probability principles and expert domain knowledge. They further depict causal relation between parent and node [44]. The probability of posterior information is dependent on prior information. Prior information contains expert knowledge and the sample data of the domain under study. The strength of Bayesian probability stands on both input data and expert information. This is because in real time scenario, not all input information is always available for the output information to depend upon. Hence, expert information helps the software community to devise conclusions. According to Bayesian theorem,

$$P(A|B) = (P(B|A) \times P(A)) \div P(B) \qquad (3)$$

P (B|A) is the conditional probability of B given A
P (A) is the prior probability of A
P (B) is the prior or marginal probability of B
P (A|B) is the conditional probability of A, given B. It is also called the posterior probability because it is derived from or depends upon the specified value of B. Using the Bayesian probability, the management can make decision on the choice of values in the parameters.

Figure 6 depicts the Bayesian nodes for our work. DI is the parent node and the DI affecting parameters are mutually independent child nodes for DI. The figure indicates that DI is influenced by all the four parameters along with skill of an inspector. However, in Bayesian Network, nodes are called variables. Hence, a Conditional Probability Table (CPT) is required to relate all the variables and parent node (parent variables).

The Conditional Probability Table (CPT) represents the uncertainty in the dependency relation. Conditional probability can be computed either through the sample data or through the expert input. As an example, the conditional probability functions for small category of projects under requirement analysis phase are given below. The Table 8 shows CPT for number of inspectors (N) and DI.

For small projects, at requirement analysis phase, If number of inspectors > 3, it is considered as level high (H). If inspectors = 3, it is considered as level moderate (M). If inspectors < 3, it is considered as level Low (L). If DI > 0.7, it is considered as level Excellent (E). If DI = 0.4 to 0.7, it is considered as level Desirable (D). If DI = 0.3 to 0.4, it is considered as level Moderate (M). If DI < 0.3, it is considered as level Poor (P). P (N= "H"| DI = "E") is the conditional probability.

Instance 1 and Instance 2 indicates that the probability of DI which is observed for the sampled five sampled projects at requirements phase under the small category of projects.
Instance 1
P(N="M"|DI="M")=P(N="M", DI = "M") / P(DI= "M") = (x / y) = (1/5)
Instance 2
P(N="M"|DI="D")=P(N="M", DI = "D") / P(DI= "D") = (x / y) =(4 / 5)

Instance 1, indicates that in requirements analysis phase for small projects, the probability of obtaining DI in the range of 0.3 to 0.4, when number of inspectors is 3 will be 20%. However, instance 2 indicates that the probability of DI to be in the range of 0.4 to 0.7 when number of inspectors is 3 will be 80%. Above instances indicate that from the input data, the managers can predict the probability of DI. DI can be predicted for all the parameters in the similar manner through input information. However, in real time scenario, due to the huge amount of data, it may not be possible to use input information to predict DI. Hence, the expert information along with sample input data is adequate to predict posterior information. Consequently, Table 9 illustrates the expert information about the desirable range of DI, IPM and inspection influencing parameters.

### 6. CONCLUSIONS

Generation of nearly a zero-defect product is one of the major issues prevailing in software industries. Effective defect management addresses the issue. Here, improved inspection positions itself to be the most significant quality assurance technique in effective defect detection and prevention.

The work is an empirical study of several projects from leading service-based and product-based software industries. Observational inferences show the capability of existing inspection technique to deliver up to 96% defect-free product. It further indicates the existence of mutually cohesive triangle comprising of process, people and effective defect management. It emphasizes upon the mutually dependent relation between process, people and effective defect management to generate high quality product. Thus, a novel and robust enhanced approach of inspection is promising to achieve effective defect management.

Enhanced inspection is achievable through a Four-Step Approach Model of Inspection (FAMI). The model comprises of i) integration of Inspection Life Cycle in V-model of software development ii) implementation of process metric, 'Depth of Inspection (DI)' to quantify the effectiveness of inspection iii) implementation of people metric, 'Inspection Performance Metric (IPM)' to measure the impact of inspection influencing parameters and iv) application of Bayesian probability enables to select an appropriate value of parameters affecting inspection to achieve desirable DI.

Table 9 specifies the range of DI, IPM values along with the required amount of inspection time, preparation time, number of inspectors and experience level of inspectors at every phase of software development to achieve effective and efficient inspection. In addition, the table also specifies the testing time, which is sufficient to achieve effective defect management in combination of specified inspection values.

Implementation of FAMI leads to the development of nearly zero-defect products. It further reflects continual process improvement in software industries through effective defect management. These are applicable for all categories of project, based on total developmental time as the mode of measurement. However, FAMI may not be applicable for projects developed with varying complexities, technology, environment and programming languages. It may not be applicable to innovative projects.

**ACKNOWLEDGEMENT**

The authors would like to acknowledge all the people from various software houses for their valuable discussions and immense help rendered in carrying out this work under the framework of non-disclosure agreement.

**BIOGRAPHY**

**Suma V.** obtained her B.E. in Information Science and Technology in 1997 and her M.S. in Software Systems in 2002. She is perusing her doctoral degree. She is a member of Research and industry incubation centre at Dayananda Sagar Institutions and works as Assistant Professor in the department of Information Science, Dayananda Sagar College of Engineering. Her main area of interest is in Software Engineering, Unified Modelling Language, Software Testing, Database Management System and Information Systems, Software Architecture.

**T. R. Gopalakrishnan Nair** holds M.Tech. (IISc. Bangalore) and Ph.D. degree in Computer Science. He has 3 decades experience in Computer Science and Engineering through research, industry and education. He has published several papers and holds patents in multi domains. He has won the PARAM Award for technology innovation. Currently he is the Director of Research and Industry in Dayananda Sagar Institutions, Bangalore, India

**Table 1.** Sampled Data from Leading Software Companies

|  | P1 | P2 | P3 | P4 | P5 | P6 | P7 | P8 | P9 | P10 | P11 | P12 | P13 | P14 | P15 |
|---|---|---|---|---|---|---|---|---|---|---|---|---|---|---|---|
| Project hours(*) | 250 | 263 | 300 | 507 | 869 | 1806 | 2110 | 4248 | 4586 | 4644 | 6944 | 7087 | 7416 | 8940 | 9220 |
| Req. Time | 25 | 23 | 32 | 55 | 72 | 73 | 800 | 1062 | 2047 | 558 | 2597 | 2237 | 2340 | 2474 | 2551 |
| Insp. Time | 3 | 3 | 4 | 6 | 6 | 7 | 48 | 107 | 200 | 36 | 281 | 225 | 235 | 234 | 250 |
| Testing time | 7 | 7 | 9 | 16 | 16 | 20 | 80 | 320 | 575 | 150 | 621 | 450 | 821 | 500 | 821 |
| Td.req | 30 | 35 | 46 | 77 | 58 | 58 | 139 | 175 | 200 | 150 | 254 | 400 | 320 | 450 | 375 |
| Ni | 16 | 17 | 31 | 40 | 19 | 28 | 69 | 80 | 77 | 40 | 112 | 175 | 156 | 200 | 175 |
| Nt | 14 | 18 | 15 | 37 | 39 | 30 | 70 | 95 | 123 | 110 | 142 | 225 | 164 | 250 | 200 |

| | | | | | | | | | | | | | | |
|---|---|---|---|---|---|---|---|---|---|---|---|---|---|---|
| Blocker | 3 | 4 | 5 | 8 | 2 | 3 | 15 | 15 | 19 | 10 | 25 | 28 | 30 | 30 | 40 |
| Critical | 4 | 4 | 6 | 9 | 2 | 3 | 14 | 26 | 25 | 25 | 30 | 40 | 37 | 65 | 42 |
| Major | 6 | 7 | 9 | 15 | 12 | 15 | 30 | 30 | 23 | 20 | 51 | 72 | 42 | 64 | 71 |
| Minor | 7 | 9 | 12 | 17 | 14 | 12 | 34 | 35 | 55 | 39 | 48 | 92 | 56 | 126 | 75 |
| Trivial | 10 | 11 | 14 | 28 | 28 | 25 | 46 | 69 | 78 | 56 | 100 | 168 | 155 | 165 | 147 |
| No. Insp | 3 | 3 | 3 | 3 | 3 | 3 | 4 | 5 | 3 | 3 | 7 | 4 | 3 | 4 | 5 |
| Prep. Time | 0.5 | 0.15 | 0.5 | 1 | 1 | 2 | 7 | 15 | 16 | 3 | 42 | 40 | 69 | 40 | 42.12 |
| Exp (**) | 1 | 1 | 1 | 2 | 5 | 5 | 3 | 5 | 2 | 5 | 7 | 6 | 3 | 6 | 3 |
| Des.Time | 46 | 40 | 46 | 110 | 110 | 167 | 400 | 1411 | 1323 | 167 | 1966 | 2820 | 2950 | 2986 | 3080 |
| Insp. Time | 6 | 4 | 5 | 11 | 16 | 20 | 48 | 143 | 128 | 16 | 200 | 220 | 300 | 250 | 345 |
| Testing time | 13 | 13 | 10 | 25 | 30 | 38 | 112 | 390 | 275 | 44 | 396 | 550 | 640 | 500 | 700 |
| Td.des | 10 | 8 | 13 | 26 | 38 | 38 | 55 | 70 | 75 | 70 | 120 | 175 | 150 | 200 | 182 |
| Ni | 5 | 3 | 6 | 14 | 16 | 19 | 24 | 34 | 33 | 28 | 77 | 80 | 86 | 90 | 78 |
| Nt | 5 | 5 | 7 | 12 | 22 | 19 | 31 | 36 | 42 | 42 | 43 | 95 | 64 | 110 | 104 |
| Blocker | 1 | 1 | 1 | 2 | 3 | 2 | 6 | 7 | 7 | 5 | 12 | 3 | 15 | 2 | 16 |
| Critical | 1 | 1 | 1 | 6 | 6 | 5 | 7 | 12 | 15 | 5 | 18 | 8 | 24 | 10 | 32 |
| Major | 2 | 2 | 2 | 5 | 3 | 10 | 12 | 15 | 15 | 13 | 25 | 45 | 36 | 30 | 44 |
| Minor | 2 | 2 | 3 | 5 | 10 | 9 | 11 | 17 | 16 | 19 | 29 | 35 | 31 | 75 | 42 |
| Trivial | 4 | 2 | 6 | 8 | 16 | 12 | 19 | 19 | 22 | 28 | 36 | 84 | 44 | 83 | 48 |
| No. Insp | 3 | 3 | 4 | 4 | 3 | 3 | 5 | 3 | 4 | 3 | 4 | 4 | 3 | 4 | 6 |
| Prep. Time | 1 | 0.5 | 1 | 1 | 2 | 2 | 7 | 25 | 25 | 2 | 33 | 50 | 61 | 60 | 123 |
| Exp (**) | 2 | 2 | 2 | 3 | 5 | 5 | 4 | 6 | 6 | 6 | 6 | 6 | 4 | 6 | 4 |
| Imp. Time | 101 | 100 | 118 | 165 | 167 | 456 | 640 | 878 | 756 | 152 | 1300 | 914 | 956 | 2134 | 2200 |
| Insp. Time | 10 | 10 | 17 | 23 | 15 | 42 | 95 | 105 | 91 | 16 | 156 | 100 | 116 | 250 | 264 |
| Testing time | 30 | 35 | 34 | 45 | 68 | 97 | 200 | 265 | 165 | 48 | 310 | 200 | 235 | 400 | 460 |
| Td.imp | 8 | 14 | 16 | 17 | 19 | 38 | 36 | 47 | 53 | 15 | 67 | 120 | 70 | 150 | 98 |
| Ni | 4 | 8 | 7 | 9 | 7 | 8 | 14 | 24 | 27 | 6 | 37 | 60 | 32 | 70 | 48 |
| Nt | 4 | 6 | 9 | 8 | 12 | 30 | 22 | 23 | 26 | 9 | 30 | 60 | 38 | 80 | 50 |
| Blocker | 1 | 2 | 2 | 1 | 0 | 0 | 3 | 5 | 4 | 2 | 5 | 2 | 7 | 3 | 10 |
| Critical | 2 | 2 | 3 | 4 | 4 | 0 | 4 | 6 | 4 | 1 | 14 | 1 | 13 | 5 | 21 |
| Major | 2 | 3 | 2 | 4 | 4 | 12 | 9 | 11 | 15 | 3 | 13 | 19 | 12 | 22 | 22 |
| Minor | 1 | 2 | 4 | 3 | 3 | 11 | 8 | 12 | 12 | 5 | 16 | 24 | 14 | 39 | 22 |
| Trivial | 2 | 5 | 5 | 5 | 8 | 15 | 12 | 13 | 18 | 4 | 19 | 74 | 24 | 81 | 23 |
| No. Insp | 3 | 3 | 3 | 3 | 3 | 3 | 5 | 5 | 5 | 3 | 3 | 4 | 6 | 4 | 4 |
| Prep. Time | 1 | 1.5 | 2 | 2 | 2 | 2 | 15 | 14 | 16 | 3 | 32 | 20 | 45 | 40 | 141 |
| Exp (**) | 2 | 2 | 2 | 3 | 5 | 5 | 5 | 6 | 6 | 5 | 7 | 6 | 4 | 6 | 5 |
| Td | 50 | 60 | 82 | 125 | 128 | 154 | 250 | 306 | 340 | 266 | 455 | 720 | 580 | 835 | 710 |
| Tc | 48 | 57 | 75 | 120 | 115 | 134 | 230 | 292 | 328 | 235 | 441 | 695 | 540 | 800 | 655 |
| Tc (%) | 96.0 | 95.0 | 91.5 | 96.0 | 89.8 | 87.0 | 92.0 | 95.4 | 96.5 | 88.3 | 96.9 | 96.5 | 93.1 | 95.8 | 92.3 |

P = Project, Req – Requirements analysis phase, Des – Design phase, Imp – Implementation phase, Insp. Time - Inspection time Td – Total number of defects, Tc – Total defects captured, Ni – Number of defects captured through inspection, Nt – Number of defects captured through testing, No. Insp – Number of inspectors, Prep.Time- Preparation time, Exp – Experience level of inspectors
(*) Total time contains documentation times, training time and release time etc. also, which are not relevant for this discussion
(**) unit of measurement for the Experience of an inspector is the number of years as an inspector

**Table 2.** Details at Requirements Phase

| Req.phase (%) | P1 | P2 | P3 | P4 | P5 | P6 | P7 | P8 | P9 | P10 | P11 | P12 | P13 | P14 | P15 |
|---|---|---|---|---|---|---|---|---|---|---|---|---|---|---|---|
| Insp. Time | 12.00 | 13.04 | 12.50 | 10.91 | 8.33 | 9.59 | 6.00 | 10.08 | 9.77 | 6.45 | 10.82 | 10.06 | 10.04 | 9.46 | 9.80 |
| Testing time | 28.00 | 30.43 | 28.13 | 29.09 | 22.22 | 27.40 | 10.00 | 30.13 | 28.09 | 26.88 | 23.91 | 20.12 | 35.09 | 20.21 | 32.18 |
| Prep. Time | 2.00 | 0.65 | 1.56 | 1.82 | 1.39 | 2.74 | 0.88 | 1.41 | 0.78 | 0.54 | 1.62 | 1.79 | 2.95 | 1.62 | 1.65 |
| Ni | 53.33 | 48.57 | 67.39 | 51.95 | 32.76 | 48.28 | 49.64 | 45.71 | 38.50 | 26.67 | 44.09 | 43.75 | 48.75 | 44.44 | 46.67 |
| Nt | 46.67 | 51.43 | 32.61 | 48.05 | 67.24 | 51.72 | 50.36 | 54.29 | 61.50 | 73.33 | 55.91 | 56.25 | 51.25 | 55.56 | 53.33 |
| Blocker | 10.00 | 11.43 | 10.87 | 10.39 | 3.45 | 5.17 | 10.79 | 8.57 | 9.50 | 6.67 | 9.84 | 7.00 | 9.38 | 6.67 | 10.67 |
| Critical | 13.33 | 11.43 | 13.04 | 11.69 | 3.45 | 5.17 | 10.07 | 14.86 | 12.50 | 16.67 | 11.81 | 10.00 | 11.56 | 14.44 | 11.20 |
| Major | 20.00 | 20.00 | 19.57 | 19.48 | 20.69 | 25.86 | 21.58 | 17.14 | 11.50 | 13.33 | 20.08 | 18.00 | 13.13 | 14.22 | 18.93 |
| Minor | 23.33 | 25.71 | 26.09 | 22.08 | 24.14 | 20.69 | 24.46 | 20.00 | 27.50 | 26.00 | 18.90 | 23.00 | 17.50 | 28.00 | 20.00 |
| Trivial | 33.33 | 31.43 | 30.43 | 36.36 | 48.28 | 43.10 | 33.09 | 39.43 | 39.00 | 37.33 | 39.37 | 42.00 | 48.44 | 36.67 | 39.20 |

**Table 3.** Details at Design Phase

| Des.phase (%) | P1 | P2 | P3 | P4 | P5 | P6 | P7 | P8 | P9 | P10 | P11 | P12 | P13 | P14 | P15 |
|---|---|---|---|---|---|---|---|---|---|---|---|---|---|---|---|
| Insp. Time | 13.04 | 10.00 | 10.87 | 10.00 | 14.55 | 11.98 | 12.00 | 10.13 | 9.67 | 9.58 | 10.17 | 7.80 | 10.17 | 8.37 | 11.20 |
| Testing time | 28.26 | 32.50 | 21.74 | 22.73 | 27.27 | 22.75 | 28.00 | 27.64 | 20.79 | 26.35 | 20.14 | 19.50 | 21.69 | 16.74 | 22.73 |
| Prep. Time | 2.17 | 1.25 | 2.17 | 0.91 | 1.82 | 1.20 | 1.75 | 1.77 | 1.89 | 1.20 | 1.68 | 1.77 | 2.07 | 2.01 | 3.99 |
| Ni | 50.00 | 57.14 | 43.75 | 52.94 | 36.84 | 21.05 | 38.89 | 51.06 | 50.94 | 40.00 | 55.22 | 50.00 | 45.71 | 46.67 | 48.98 |
| Nt | 50.00 | 62.50 | 53.85 | 46.15 | 57.89 | 50.00 | 56.36 | 51.43 | 56.00 | 60.00 | 35.83 | 54.29 | 42.67 | 55.00 | 57.14 |
| Blocker | 10.00 | 12.50 | 7.69 | 7.69 | 7.89 | 5.26 | 10.91 | 10.00 | 9.33 | 7.14 | 10.00 | 1.71 | 10.00 | 1.00 | 8.79 |
| Critical | 10.00 | 12.50 | 7.69 | 23.08 | 15.79 | 13.16 | 12.73 | 17.14 | 20.00 | 7.14 | 15.00 | 4.57 | 16.00 | 5.00 | 17.58 |
| Major | 20.00 | 25.00 | 15.38 | 19.23 | 7.89 | 26.32 | 21.82 | 21.43 | 20.00 | 18.57 | 20.83 | 25.71 | 24.00 | 15.00 | 24.18 |
| Minor | 20.00 | 25.00 | 23.08 | 19.23 | 26.32 | 23.68 | 20.00 | 24.29 | 21.33 | 27.14 | 24.17 | 20.00 | 20.67 | 37.50 | 23.08 |
| Trivial | 40.00 | 25.00 | 46.15 | 30.77 | 42.11 | 31.58 | 34.55 | 27.14 | 29.33 | 40.00 | 30.00 | 48.00 | 29.33 | 41.50 | 26.37 |

**Table 4.** Details at Implementation Phase

| Imp. Phase (%) | P1 | P2 | P3 | P4 | P5 | P6 | P7 | P8 | P9 | P10 | P11 | P12 | P13 | P14 | P15 |
|---|---|---|---|---|---|---|---|---|---|---|---|---|---|---|---|
| Insp. Time | 9.90 | 10.00 | 14.41 | 13.94 | 8.98 | 9.21 | 14.84 | 11.96 | 12.04 | 10.53 | 12.00 | 10.94 | 12.13 | 11.72 | 12.00 |
| Testing time | 29.70 | 35.00 | 28.81 | 27.27 | 40.72 | 21.27 | 31.25 | 30.18 | 21.83 | 31.58 | 23.85 | 21.88 | 24.58 | 18.74 | 20.91 |
| Prep. Time | 0.99 | 1.50 | 1.69 | 1.21 | 1.20 | 0.44 | 2.34 | 1.59 | 2.12 | 1.97 | 2.46 | 2.19 | 4.71 | 1.87 | 6.41 |
| Ni | 50.00 | 57.14 | 43.75 | 52.94 | 36.84 | 21.05 | 38.89 | 51.06 | 50.94 | 40.00 | 55.22 | 50.00 | 45.71 | 46.67 | 48.98 |
| Nt | 50.00 | 42.86 | 56.25 | 47.06 | 63.16 | 78.95 | 61.11 | 48.94 | 49.06 | 60.00 | 44.78 | 50.00 | 54.29 | 53.33 | 51.02 |
| Blocker | 12.50 | 14.29 | 12.50 | 5.88 | 0.00 | 0.00 | 8.33 | 10.64 | 7.55 | 13.33 | 7.46 | 1.67 | 10.00 | 2.00 | 10.20 |
| Critical | 25.00 | 14.29 | 18.75 | 23.53 | 21.05 | 0.00 | 11.11 | 12.77 | 7.55 | 6.67 | 20.90 | 0.83 | 18.57 | 3.33 | 21.43 |
| Major | 25.00 | 21.43 | 12.50 | 23.53 | 21.05 | 31.58 | 25.00 | 23.40 | 28.30 | 20.00 | 19.40 | 15.83 | 17.14 | 14.67 | 22.45 |
| Minor | 12.50 | 14.29 | 25.00 | 17.65 | 15.79 | 28.95 | 22.22 | 25.53 | 22.64 | 33.33 | 23.88 | 20.00 | 20.00 | 26.00 | 22.45 |
| Trivial | 25.00 | 35.71 | 31.25 | 29.41 | 42.11 | 39.47 | 33.33 | 27.66 | 33.96 | 26.67 | 28.36 | 61.67 | 34.29 | 54.00 | 23.47 |

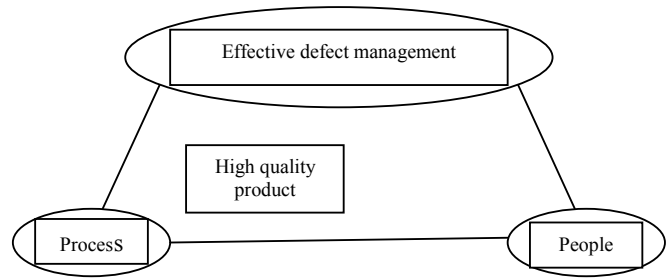

**Figure 1.** Existence of Mutually Cohesive Triangle

**Table 5.** Percentage of Occurrences of Various Types of Defects at Each Phase of Software Development for the Undertaken Categories of Projects

| Phase | Defect type (%) | Small projects | Medium projects | Large projects | Average |
|---|---|---|---|---|---|
| Req | Blocker | 0% to 10% | 0% to 10% | 0% to 10% | 5% to 10% |
| | Critical | 0% to 15% | 0% to 15% | 0% to 15% | 10% to 15% |
| | Major | 10% to 20% | 20% to 30% | 10% to 20% | 10% to 20% |
| | Minor | 20% to 25% | 20% to 30% | 15% to 30% | 20% to 25% |
| | Trivial | 30% to 50% | 30% to 45% | 30% to 50% | 30% to 40% |
| Des | Blocker | 0% to 15% | 0% to 10% | 0% to 10% | 0% to 10% |
| | Critical | 5% to 15% | 5% to 15% | 5% to 20% | 5% to 17% |
| | Major | 5% to 25% | 15% to 25% | 15% to 25% | 10% to 20% |
| | Minor | 15% to 25% | 20% to 30% | 20% to 30% | 15% to 25% |
| | Trivial | 25% to 45% | 25% to 45% | 25% to 50% | 25% to 40% |
| Imp | Blocker | 0% to 10% | 0% to 10% | 0% to 10% | 0% to 10% |
| | Critical | 10% to 25% | 0% to 15% | 0% to 15% | 0% to 20% |
| | Major | 10% to 25% | 20% to 35% | 10% to 25% | 10% to 25% |
| | Minor | 10% to 25% | 20% to 35% | 20% to 35% | 10% to 25% |
| | Trivial | 25% to 45% | 25% to 45% | 25% to 55% | 25% to 40% |

**Table 6.** DI Estimation

|  | P1 | P2 | P3 | P4 | P5 | P6 | P7 | P8 | P9 | P10 | P11 | P12 | P13 | P14 | P15 |
|---|---|---|---|---|---|---|---|---|---|---|---|---|---|---|---|
| Project hours(*) | 250 | 263 | 300 | 507 | 869 | 1806 | 2110 | 4248 | 4586 | 4644 | 6944 | 7087 | 7416 | 8940 | 9220 |
| DI at req. phase | 0.53 | 0.49 | 0.67 | 0.52 | 0.33 | 0.48 | 0.50 | 0.46 | 0.39 | 0.27 | 0.44 | 0.44 | 0.49 | 0.44 | 0.47 |
| DI at des. phase | 0.50 | 0.38 | 0.46 | 0.54 | 0.42 | 0.50 | 0.44 | 0.49 | 0.44 | 0.40 | 0.64 | 0.46 | 0.57 | 0.45 | 0.43 |
| DI at imp. phase | 0.50 | 0.57 | 0.44 | 0.53 | 0.37 | 0.21 | 0.39 | 0.51 | 0.51 | 0.40 | 0.55 | 0.50 | 0.46 | 0.47 | 0.49 |
| Td | 50 | 60 | 82 | 125 | 128 | 154 | 250 | 306 | 340 | 266 | 455 | 720 | 580 | 835 | 710 |
| Tc | 48 | 57 | 75 | 120 | 115 | 134 | 230 | 292 | 328 | 235 | 441 | 695 | 540 | 800 | 655 |
| Tc (%) | 96.00 | 95.00 | 91.46 | 96.00 | 89.84 | 87.01 | 92.00 | 95.42 | 96.47 | 88.35 | 96.92 | 96.53 | 93.10 | 95.81 | 92.25 |

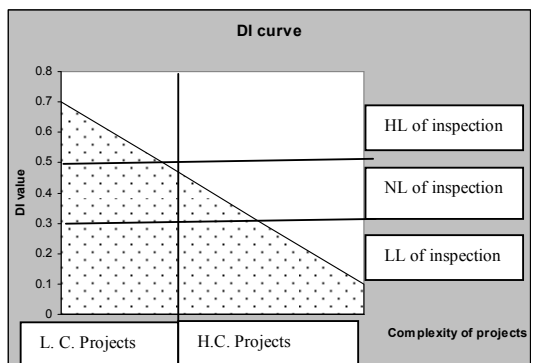

HL = High level; NL = Normal level; LL = Low level;
L.C.= Lesser complexity; H.C. = Higher complexity

**Figure 4.** Variation of Depth of Inspection with Increase in Complexity of the Projects

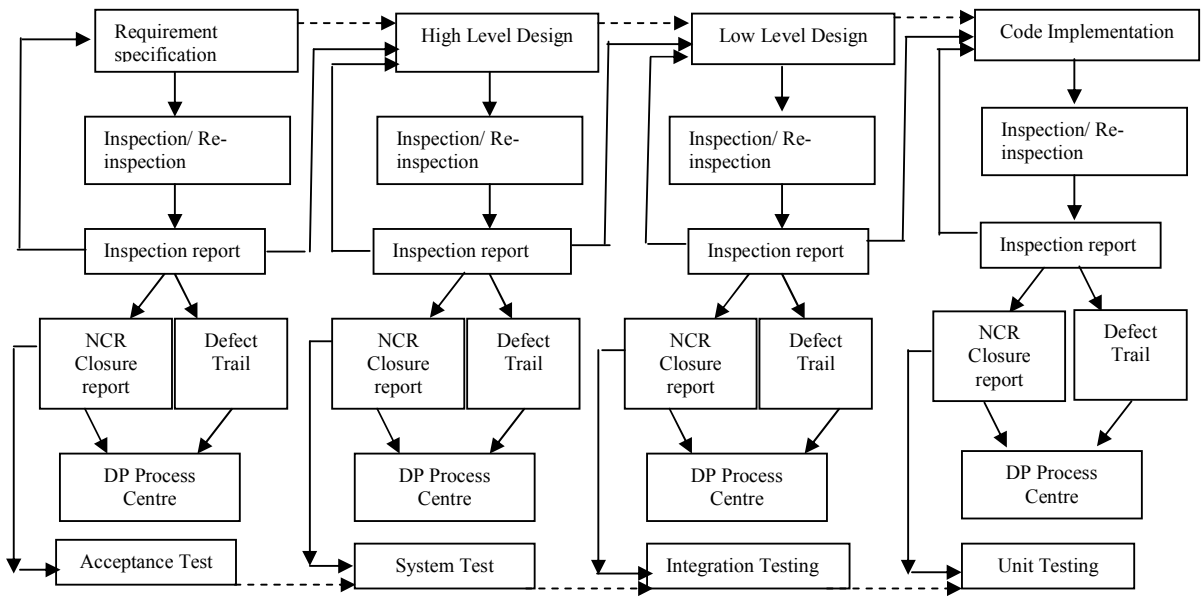

**Figure 2.** Integration of Inspection Life Cycle in V-Model

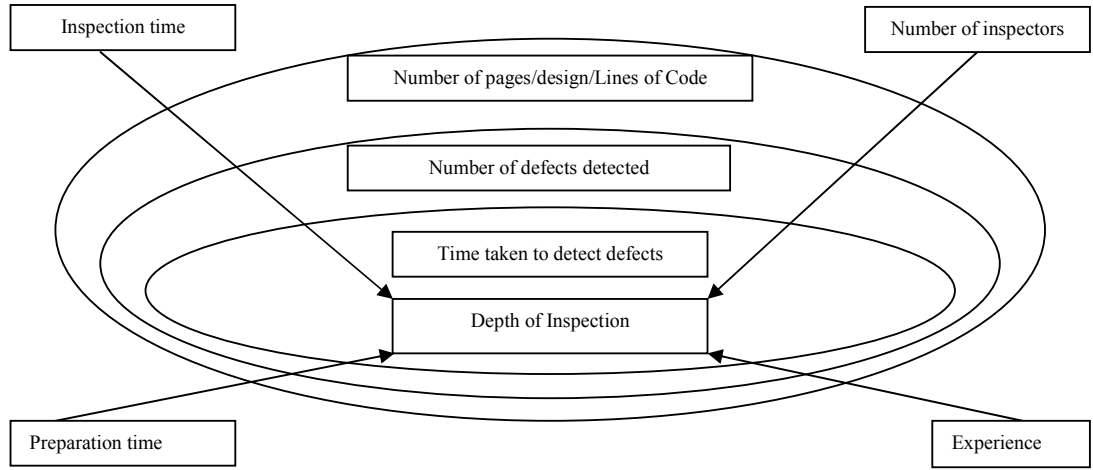

**Figure 5.** Influence of Various Parameters on Depth of Inspection

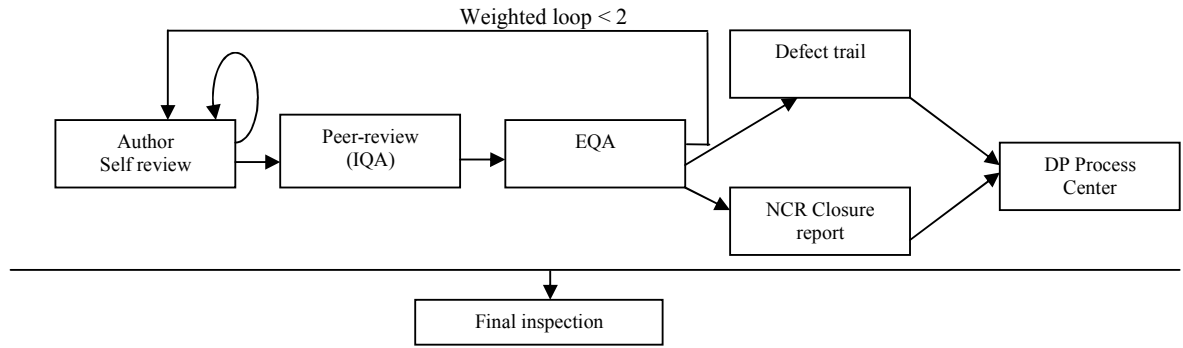

**Figure 3.** Inspection Process for Deliverable

**Table 7.** Impact Analysis of DI and IPM on the Sampled Projects

| Phase | | P1 | P2 | P3 | P4 | P5 | P6 | P7 | P8 | P9 | P10 | P11 | P12 | P13 | P14 | P15 |
|---|---|---|---|---|---|---|---|---|---|---|---|---|---|---|---|---|
| Req | DI | 0.53 | 0.49 | 0.67 | 0.52 | 0.33 | 0.48 | 0.5 | 0.46 | 0.39 | 0.27 | 0.44 | 0.44 | 0.49 | 0.44 | 0.47 |
| | IPM | 1.52 | 1.8 | 2.3 | 1.9 | 0.9 | 1.04 | 0.31 | 0.13 | 0.12 | 0.34 | 0.05 | 0.17 | 0.17 | 0.18 | 0.12 |
| | Exp | 1 | 1 | 1 | 2 | 5 | 5 | 3 | 5 | 5 | 2 | 7 | 6 | 3 | 6 | 3 |
| Des | DI | 0.5 | 0.38 | 0.46 | 0.54 | 0.42 | 0.5 | 0.44 | 0.49 | 0.44 | 0.4 | 0.64 | 0.46 | 0.57 | 0.45 | 0.43 |
| | IPM | 0.24 | 0.22 | 0.25 | 0.29 | 0.3 | 0.29 | 0.09 | 0.07 | 0.05 | 0.52 | 0.08 | 0.07 | 0.08 | 0.07 | 0.03 |
| | Exp | 2 | 2 | 2 | 3 | 5 | 5 | 4 | 6 | 6 | 6 | 6 | 6 | 4 | 6 | 4 |
| Imp | DI | 0.5 | 0.57 | 0.44 | 0.53 | 0.37 | 0.21 | 0.39 | 0.51 | 0.51 | 0.4 | 0.55 | 0.5 | 0.46 | 0.47 | 0.49 |
| | IPM | 0.12 | 0.23 | 0.12 | 0.12 | 0.14 | 0.06 | 0.03 | 0.04 | 0.05 | 0.11 | 0.07 | 0.13 | 0.03 | 0.06 | 0.03 |
| | Exp | 2 | 2 | 2 | 3 | 5 | 5 | 5 | 6 | 5 | 6 | 7 | 6 | 4 | 6 | 5 |
| Total project | Td | 50 | 60 | 82 | 125 | 128 | 154 | 250 | 306 | 340 | 266 | 455 | 720 | 580 | 835 | 710 |
| | Tc | 48 | 57 | 75 | 120 | 115 | 134 | 230 | 292 | 328 | 235 | 441 | 695 | 540 | 800 | 655 |
| | Tc (%) | 96 | 95 | 91.46 | 96 | 89.84 | 87.01 | 92 | 95.42 | 96.47 | 88.35 | 96.92 | 96.53 | 93.1 | 95.81 | 92.25 |

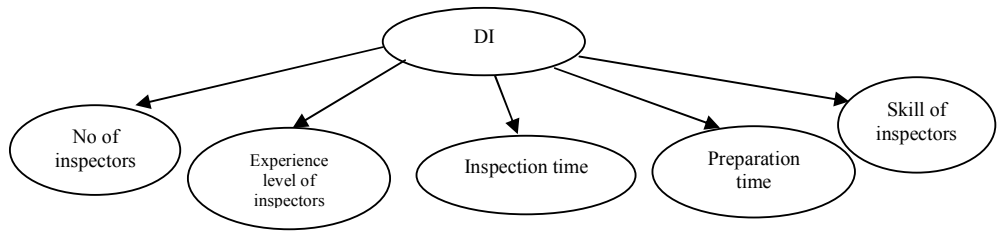

**Figure 6.** Causal Relations Existing between Bayesian Nodes for Depth of Inspection and its Influencing Parameters

**Table 8.** CPT for Four Parameters and DI

|  | Poor (P) | Moderate (M) | Desirable (D) | Excellent (E) |
|---|---|---|---|---|
| Low (L) | P( N= "L"\|DI = L") | P( N= "L"\|DI = "M") | P( N= "L"\|DI = "D") | P( N= "L"\|DI = "E") |
| Moderate (M) | P( N= "M"\|DI = "L") | P( N= "M"\|DI = "M") | P( N= "M"\|DI = "D") | P( N= "M"\|DI = "E") |
| High (H) | P( N= "H"\|DI = "L") | P( N= "H"\|DI = "M") | P( N= "H"\|DI = "D") | P( N= "H"\|DI = "E") |

**Table 9.** Desired Ranges for the Parameters Influencing Depth of Inspection

| Phase | Parameters | Small project | Medium project | Large project |
|---|---|---|---|---|
| Req | DI | 0.4 – 0.7 | 0.4 – 0.7 | 0.4 – 0.7 |
|  | IPM | 1-2.5 | 0.1 – 1 | Above 0.05 |
|  | Inspection time | 10% -15% | 10% -15% | 10% -15% |
|  | Preparation time | 10% -20% | 10% -20% | 10% -20% |
|  | No of inspectors | 3 | 3 - 5 | Above 4 |
|  | Exp of inspectors | 1 -3 | 3 – 5 | Above 3 |
|  | Testing time | 20% - 30% | 20% - 35% | 20% - 35% |
| Des | DI | 0.4 – 0.7 | 0.4 – 0.7 | 0.4 – 0.7 |
|  | IPM | Above 0.1 | Above 0. 1 | Above 0.05 |
|  | Inspection time | 10% -15% | 10% -15% | 10% -15% |
|  | Preparation time | 10% -20% | 10% -20% | 10% -20% |
|  | No of inspectors | 4 | 3 - 5 | Above 4 |
|  | Exp of inspectors | 2 - 3 | Above 4 | Above 4 |
|  | Testing time | 20% - 30% | 20% - 35% | 20% - 35% |
| Imp | DI | 0.4 – 0.7 | 0.4 – 0.7 | 0.4 – 0.7 |
|  | IPM | Above 0.1 | Above 0.05 | Above 0.05 |
|  | Inspection time | 10% -15% | 10% -15% | 10% -15% |
|  | Preparation time | 10% -20% | 10% -20% | 10% -20% |
|  | No of inspectors | 3 | 3 – 5 | Above 4 |
|  | Exp of inspectors | 2 – 3 | Above 4 | Above 4 |
|  | Testing time | 20% - 30% | 20% - 35% | 20% - 35% |